\documentstyle[12pt,epsf]{article}

\advance\voffset by -1.5cm
\advance\hoffset by -1.5cm
\def\be{\begin{equation}}
\def\ee{\end{equation}}

\def\pmb#1{\setbox0=\hbox{#1}
 \kern-.025em\copy0\kern-\wd0
 \kern.05em\copy0\kern-\wd0
 \kern-.025em\raise.0433em\box0 }

\def\3{\ss}
\def\sq{\hbox{\rlap{$\sqcap$}$\sqcup$}}
\def\qed{\ifmmode\sq\else{\unskip\nobreak\hfil
\penalty50\hskip1em\null\nobreak\hfil\sq
\parfillskip=0pt\finalhyphendemerits=0\endgraf}\fi}

\def\bbbz {{\sf Z\!\!Z}}

\def\bbbr {{\rm I\!R}}

\def\bq{{\bf q}}
\def\bp{{\bf p}}

\def\bv{{\bf v}}
\def\bbeta{\mbox{\boldmath{$\beta$}}}

\def\bPhi{\mbox{\boldmath{$\Phi$}}}

\def\ss{\bf S}

\newcommand{\bea}{\begin{eqnarray}}
\newcommand{\eea}{\end{eqnarray}}
\newcommand{\pq}{$(p,q)$ }

\def\href#1#2{#2}

\def\IZ{\relax\ifmmode\mathchoice
{\hbox{\cmss Z\kern-.4em Z}}{\hbox{\cmss Z\kern-.4em Z}}
{\lower.9pt\hbox{\cmsss Z\kern-.4em Z}}
{\lower1.2pt\hbox{\cmsss Z\kern-.4em Z}}\else{\cmss Z\kern-.4em
   Z}\fi}
\def\IR{\relax{\rm I\kern-.18em R}}

\font\cmss=cmss10 \font\cmsss=cmss10 at 7pt

\begin{document}

\thispagestyle{empty}
\def\thefootnote{\fnsymbol{footnote}}
\begin{flushright}
  hep-th/9804160\\
  HUTP-98/A035\\
  SU-ITP-98-14
 \end{flushright}

\begin{center}\LARGE
{\bf String Webs and 1/4 BPS Monopoles}
\end{center}
\vskip 0.5cm
\begin{center}
{\large  Oren Bergman\footnote{E-mail  address: {\tt
bergman@string.harvard.edu}}}

\vskip 0.2 cm
{\it Lyman Laboratory of Physics\\
Harvard University\\
Cambridge, MA 02138}

\vskip 0.5 cm
{\large Barak Kol
\footnote{E-mail  address:
{\tt barak@leland.stanford.edu}}}

\vskip 0.2 cm
{\it Department of Physics\\
Stanford University\\
Stanford, CA 94305}

\end{center}

\vskip 0.5cm

\begin{center}
April 1998
\end{center}

\vskip 0.5cm

\begin{abstract}
We argue for the existence of many new 1/4 BPS states in $N=4$ $SU(N_c)$
Super-Yang-Mills theory with $N_c\geq 3$, by constructing them from
supersymmetric string webs whose external strings terminate on parallel
D3-branes. The masses of the string webs are shown to agree with the BPS bound
for the corresponding states in SYM. We identify the curves of marginal
stability, at which these states decay into other BPS states. We find the
bosonic and fermionic zero modes of the string webs, and thereby the degeneracy
and spin content of some of the BPS states. States of arbitrarily high spin are
predicted in this manner, all of which become massless at the conformal point.
For $N_c\geq 4$ we find BPS states which transform in long multiplets, and are
therefore not protected against becoming stable non-BPS states as moduli are
varied. The mass of these extremal non-BPS states is constrained as they are
connected to BPS states. Analogous geometric phenomena are anticipated.
\end{abstract}

\vfill
\begin{center}
PACS codes: 11.25.-w, 11.15.-q, 11.30.Pb
\end{center}

\setcounter{footnote}{0}
\def\thefootnote{\arabic{footnote}}
\newpage

\renewcommand{\theequation}{\thesection.\arabic
{equation}}

\begin{flushright}
To the memory of Alon Cohen (1968-1997)\\BK
\end{flushright}

\section{Introduction and summary}

\noindent String theory has proven to be a powerful tool
for studying various non-perturbative aspects of
supersymmetric quantum field theories. D-branes have
played a central role in this process, both by providing
non-perturbative states in string theories (whose low
energy limits are field theories)
which are required by the various string dualities
\cite{polchinski}, and by providing a new test bed
for quantum field theories through their
world-volume dynamics \cite{giveon_review}.
D-strings in particular provide yet another insight
into supersymmetric field theories, namely in determining
the BPS spectrum thereof. For example, a D-string which
ends on a D3-brane corresponds to a BPS monopole in
the world-volume $N=4$ $U(N_c)$ SYM theory,
where $N_c$ is the number of parallel D3-branes
\cite{tgg}.
Similarly a bound
state of $p$ fundamental strings and $q$ D-strings,
where $p$ and $q$ are co-prime integers, corresponds
to a $(p,q)$ BPS dyon. Such states are predicted by
the conjectured $SL(2,\bbbz)$ symmetry of $N=4$
SYM \cite{Sen_dyon},
but their existence from the field theory point of view
has only been established in a limited number of cases
\cite{dyons}. In the string theory approach their existence
follows directly from the existence of $(p,q)$ strings
in type IIB string theory \cite{Schwarz_pq,wittenbound}.
By studying a D3-brane probe near an orientifold 7-plane,
whose quantum resolution consists of two mutually non-local
7-branes \cite{sen_f}, one can similarly construct
the monopole and dyon of Seiberg-Witten theory, {\it i.e.}
$N=2$ $SU(2)$ SYM, as appropriate $(p,q)$ strings between
the D3-brane and either 7-brane \cite{sen_bps}.

In addition to $(p,q)$ strings, type IIB string theory contains more general
stringy objects consisting of several connected $(p,q)$ strings, or ``string
webs'' [10--19]\footnote{ Analogous \pq 5-brane webs were constructed in
\cite{AH,AHK}, and were used to study five-dimensional $N=1$ theories.}. The
simplest example is a fundamental string, {\it i.e.} $(1,0)$, which terminates
on a D-string, {\it i.e.} $(0,1)$, to form a ``three-string junction''. By
charge conservation the third string must carry the charges $(1,1)$.
Furthermore, unbroken supersymmetry requires the configuration to be planar, and
the relative angles of the three strings to be given by $\mbox{Arg}(p_i +
q_i\tau)$, where $\tau = i/g_s + a$, $g_s$ is the string coupling constant, and
$a$ is the expectation value of the RR scalar field. Under these conditions
string webs preserve 8 of the 32 supersymmetries of type IIB string theory, and
are classically static. 

Just as strings ending on D-branes correspond to BPS
states in the world-volume field theories, string webs whose external strings
end on D-branes should correspond to BPS states, as long as the whole
configuration preserves some supersymmetry. This idea has been successfully
applied to understanding exceptional gauge symmetry enhancement in eight
dimensions \cite{gz}, finding the BPS spectrum of 5d $N=1$ gauge theories
\cite{AHK,KolRahmfeld}, constructing 1/4 BPS states in 4d $N=4$ $SU(3)$ SYM
\cite{Bergman}, deriving the entire BPS spectrum of Seiberg-Witten theory in the
D3-brane probe picture \cite{bf,mns,bf2}, and producing $E_8$ flavor multiplets
of quarks in theories with exceptional global symmetry \cite{imamura}.

In this paper we are interested in the application
of general string webs to the construction of 1/4 BPS
states in $N=4$ $SU(N_c)$ SYM. In particular,
we shall generalize the results of \cite{Bergman}
for arbitrary $N_c$ and arbitrary charges.
In addition, we shall determine the
fermionic zero modes of the string webs,
and thus the degeneracy of
some
of the BPS states.
This will be illustrated with some simple examples.

Since any \pq string can end on a D3-brane, it follows that string webs can end
on D3-branes, and therefore give rise to new states in the world-volume field
theory when the number of D3-branes is greater than two. The addition of
D3-branes further limits the number of unbroken supersymmetries to 4, so the
string webs should correspond to 1/4 BPS states in $N=4$ $SU(N_c)$ SYM. Such
states can only appear for $N_c\geq 3$, which is consistent with the requirement
that there be at least three D3-branes to end on. Indeed in \cite{Bergman} it
was shown that the 3-string junction corresponds to a 1/4 BPS state in $SU(3)$
SYM carrying the charges $((1,0),(0,1))$. The mass of the 3-string junction
reproduces the BPS bound of $N=4$ SYM. Furthermore, it was shown that this state
becomes marginally stable on a certain closed curve
in moduli space, and that it
in fact decays as the curve is crossed.

The generalization to arbitrary $N_c$ and arbitrary charges follows by
considering more general string webs, which include more external strings, as
well as internal strings. These also include degenerate string webs, in which
some external strings that carry identical charges (and are therefore parallel)
overlap. In this case the overlapping strings can end on the same D3-brane, and
thereby give rise to non-co-prime world-volume charges. Therefore unlike 1/2 BPS
states, the electric and magnetic charge pairs of
1/4 BPS states need {\it not}
be relatively prime. The only requirement is that the different
charge pairs not have a common divisor. In that case the string web would be
{\it reducible}, and therefore
at most marginally stable against decay into
a multi-particle state. Another kind of degeneration occurs when
an external string shrinks, {\it i.e.} when one of the D3-branes coincides with
a junction point. For the simple 3-string junction studied in \cite{Bergman}
this is precisely where the state becomes
marginal, and decays into the two states corresponding to the two
remaining strings.
The decay corresponds to separating the two strings along the
common D3-brane.
The same holds in the more general case; the state
corresponding to the original string web becomes marginal, and decays into two
other states, which correspond to the two remaining
webs sharing the D3-brane in question.

The degeneracy of a BPS state which is represented by a string web, and
therefore the spins of its components, is determined by the zero modes of the
web \cite{KolRahmfeld}.
The bosonic zero modes (BZM) come in two varieties, {\it local}
and {\it global}. Local BZM keep the external strings fixed, whereas global BZM
move them. The global BZM divide into planar and transverse modes with respect
to the web, where the transverse modes correspond to translations transverse to
the web, and the planar modes include the two planar translations as well as
other modes. The local BZM, on the other hand, are purely planar. In the
presence of D3-branes all the global planar modes and 4 of the global transverse
modes are eliminated, leaving only 3 transverse modes corresponding to
translations along the D3-branes, and the local BZM. The latter correspond to
loops of internal strings, {\it i.e.} internal faces, which can shrink and
expand without changing the mass of the web \cite{AHK}.
We note that in geometrical lifts to M theory each 
local BZM is
expected to complexify
\cite{Witten_phase,BK5vM,LV}.

To find the fermionic zero modes (FZM) one must solve the massless Dirac
equation on the web. This is done by gluing together the solutions on the
individual strings, {\it i.e.} constant 16 component real spinors, at the
junctions and boundaries of the web. The fermionic junction and boundary
constraints are found by applying the {\it locally} preserved supersymmetries to
the bosonic ones. Note that although there are only 4 {\it globally} preserved
supersymmetries, each junction (or boundary) actually preserves 8 {\it locally}.
For a string web with $F_{int}$ internal faces and $E_{ext}$ external strings we
find $n_{FZM}=8F_{int}+4E_{ext}$. This is shown both by solving the FZM
equations, and by generating FZM from bosonic configurations by applying
different supersymmetries. The result includes FZM which were previously
ignored.

Upon lifting the web to an M-theory membrane, $g=F_{int}$ is the genus,
$b=E_{ext}$ is the number of boundaries, and we have $n_{FZM}=4(2-\chi)$, where
$\chi$ is the Euler character of the membrane. Using the same methods we find
that for an unbounded web $n_{FZM}=8F_{int}+8E_{ext}$, while for a periodic web
\cite{Sen_net} $n_{FZM}=8(F+2)$, where $F$ is the number of faces in a unit cell.

To compute the degeneracy of the state one needs to solve for the ground state
of the supersymmetric quantum mechanics defined by the zero modes. For webs with
no (local) BZM the quantization is straightforward, and we indeed get
predictions for the
degeneracy of the corresponding monopoles, and therefore for the highest spin in
the supermultiplet. In the general case, even without the solution at hand, we
expect that the degeneracy increases with the number of fermionic zero modes,
as the spin is bounded by $\left| j_{max} \right| \leq n_{FZM}/8
= E_{ext}/2 + F_{int}$ (the bound is saturated for $F_{int}=0$).
Since there exist webs with an arbitrary number of internal faces, there are 1/4
BPS states of arbitrary degeneracy, and therefore particles of arbitrarily high
spin. All these particles will become massless when three (or more) D3-branes
coincide. Indeed, having a tower of states with arbitrarily high spin becoming
massless is a signature of a conformal field theory.

For $N_c \geq 4$ we encounter an interesting effect. As long as the D3-branes
lie in a single plane, and marginal stability curves are not crossed, there
exist planar webs connecting them, and therefore the states are BPS. On the
other hand, planar D3-brane configurations are only a subspace of real
co-dimension $4(N_c-3)$ of the moduli space. If we move out of this subspace the
web can no longer be planar, and therefore the state cannot remain
supersymmetric. BPS states thus evolve smoothly into non-BPS states as the
moduli are varied. Alternatively, a non-BPS state may become BPS on a subspace
of the moduli space.
Non-BPS states have at least 16 FZM, and therefore transform in 
{\it long}
multiplets. 
This would avoid
contradiction with \cite{OliveWitten}, if the BPS states to
which they evolve would transform in long, rather than short or medium,
multiplets
\footnote{We thank E. Witten for suggesting the solution.}.
Indeed
the BPS configuration
is the endpoint of non-BPS configurations, and
must therefore possess
at least
as many fermionic zero modes. Our formula for the number of FZM
confirms this. We shall refer to such BPS states as ``long BPS states" or
``accidental BPS states"
as they exist only on a submanifold of moduli space.
Despite being long, the mass of these states is protected against quantum
corrections since they still preserve 4 supersymmetries.

The non-BPS state corresponding to the non-planar web will be
extremal, and therefore stable
\footnote{Other examples of stable non-BPS states
were discussed by Sen \cite{Sen-nonBPS}.}. 
This is realized classically as a
zero force condition at each junction. Since different junctions will in general
lie in different planes however, 
the configuration breaks supersymmetry. The classical
mass of the non-BPS state can be seen to be strictly larger than the BPS bound.
Furthermore, one expects quantum corrections to this mass. Since these states
are continuously connected to BPS states, the quantum corrections may be studied
by expanding their mass about the BPS bound as a function of parameters which
measure the deviation from planarity.

It should be clarified that throughout this paper we study the string web model
without performing a full quantum field theory analysis. We do not check whether
the states described by the web exist as solutions of $N=4$ SYM, nor that they
cannot exist beyond marginal stability. 
Semiclassical methods may probe these
questions at weak coupling, where a recently found 
classical solution \cite{HHS}
may be used as a starting point.
Similar comments hold for the other results as well.

Let us point out some open questions:
\begin{itemize}
\item{A quantum field theory analysis, perhaps using semiclassical
methods, is required
 to check the phenomena that we describe at the string level.}
\item{The stable non-BPS states are expected to receive quantum
corrections.
 Being continuously connected to BPS states might facilitate
such a computation.}
\item{We found the bosonic and fermionic zero modes of a monopole
represented by a
web. In the general case, it remains to find the ground
states of the
appropriate quantum mechanics.}
\item{All the discussed phenomena should have a geometric analog.
For example,
 the $4d$ $N=4$ theory can be realized as type II on
$K3 \times T^2$. The $K3$
should have an $A_{N_c-1}$ singularity to decouple gravity,
and provide
the $SU(N_c)$ gauge group (as in geometric engineering,
for example \cite{GeoEng}).
Particles would correspond to some submanifolds.
The $1/4$ BPS states would correspond to some special submanifolds.
Marginal
stability will indicate that these special
submanifolds can disappear beyond some walls in moduli space, as
long as they become reducible. Accidental BPS states would
correspond to special submanifolds that
exist only in subspaces of moduli space,
and would be continuously
connected to non-special minimal volume submanifolds.}
\end{itemize}

The paper is organized as follows. In section~2 we review string webs and their
dual grid diagrams. We also discuss degenerations, bosonic zero modes, and
classification of string webs. In section~3 we generalize the results of
\cite{Bergman} to arbitrary $N_c$ and arbitrary charges.
We also obtain the fermionic zero modes, and discuss the classification of BPS states
in $N=4$ SYM.
In section~4 we discuss the smooth connection between
long BPS and
stable non-BPS states for $N_c\geq 4$.

\section{Review of string webs and grid diagrams}
\setcounter{equation}{0}

\noindent In this section we shall review the
construction and properties of multi-string webs
and their dual grid diagrams.
Most of what follows applies equally well to 5-brane webs
\cite{AH,AHK}.
We shall use standard notation
to denote the web variables: $V$ for the number
of vertices, $E$ for the number of edges, and $F$
for the number of faces.
These satisfy the well known Euler relation
\be
 V - E + F = 1 \; .
\label{euler}
\ee
The webs in question contain both external and internal edges, and therefore
both external and internal faces. We shall denote these by $E_{ext}, E_{int}$
and $F_{ext}, F_{int}$, respectively, where $E_{ext}+E_{int}=E$ and
$F_{ext}+F_{int}=F$. It is easy to see that
\be
E_{ext}=F_{ext} \; .
\label{externals}
\ee
We shall assume that generically all vertices consist of
3 edges. This gives another relation
\be
3V = E_{ext} + 2 E_{int} \; .
\label{3junction}
\ee
If more than 3 edges meet at a point we shall
interpret this as a {\it degeneration} of the web, in which
some internal edges have a vanishing length.

\subsection{string webs}

\noindent A type IIB fundamental string which ends on a
D-string forms a three-legged object known as a 3-string
junction \cite{ASY,SchwarzTASI}. By charge conservation,
the third leg must carry
the charges of both the fundamental and the Dirichlet
strings, and therefore
corresponds to a $(1,1)$ string. The condition of
unbroken supersymmetry
restricts the configuration to be planar, and the
relative directions of the
three strings in the $(x,y)$ plane to be given by
\be
 e^{i\theta_i} \propto \Delta x_i+i \Delta y_i
   \propto p_i + q_i\tau \; ,
\label{angles}
\ee
where $\tau = i/g_s + a$, $g_s$ is the type IIB string coupling,
and $a$ is the expectation value of the RR scalar field.
Since the tension of a
$(p,q)$ string is given (in the Einstein frame) by
\be
T_{(p,q)} = {1\over 2\pi\alpha'\sqrt{g_s}}|p + q\tau| \; ,
\ee
the above supersymmetry condition implies that the
force due to
the three string tensions vanishes, consistent with the
idea that BPS states
are stable.

Just as the fundamental string is but a single component of
an infinite
$SL(2,\bbbz)$ multiplet of $(p,q)$ strings \cite{Schwarz_pq},
the above
configuration is only one component of a multiplet of 3-string
junctions, which
are all related by $SL(2,\bbbz)$. All the components satisfy
charge
conservation,
\be
 \sum_{i=1}^3 p_i = \sum_{i=1}^3 q_i = 0 \; ,
\label{chargesum}
\ee
as well as the supersymmetry condition (\ref{angles}).
These are the
simplest examples of string webs, and in fact constitute
the building blocks for all string webs.

We define an {\it irreducible} string web as a collection of $(p,q)$ strings
connected by 3-string junctions, with no disconnected components, such that the
$i$'th string lies along $p_i+q_i\tau$. The actual orientation of the string is
determined up to a change in the overall sign of $(p_i,q_i)$. The external
strings are taken to be infinite, and oriented outward. The supersymmetries left
unbroken by a string web oriented in the $(x^8,x^9)$ plane
are given by
$\epsilon_LQ_L +
\epsilon_RQ_R$, with
\be
 \epsilon_L = \Gamma_0\Gamma_9\epsilon_L \;, \qquad
 \epsilon_R = - \Gamma_0\Gamma_9\epsilon_R \;, \qquad
 \epsilon_L = \Gamma_0\Gamma_8\epsilon_R \; .
\ee
Thus string webs are invariant under 1/4 of the original
supersymmetry
of type IIB string theory, {\it i.e.} 8 supercharges.
For some simple examples see figure~1.
\begin{figure}[htb]
\epsfxsize= 4in
\centerline{\epsffile{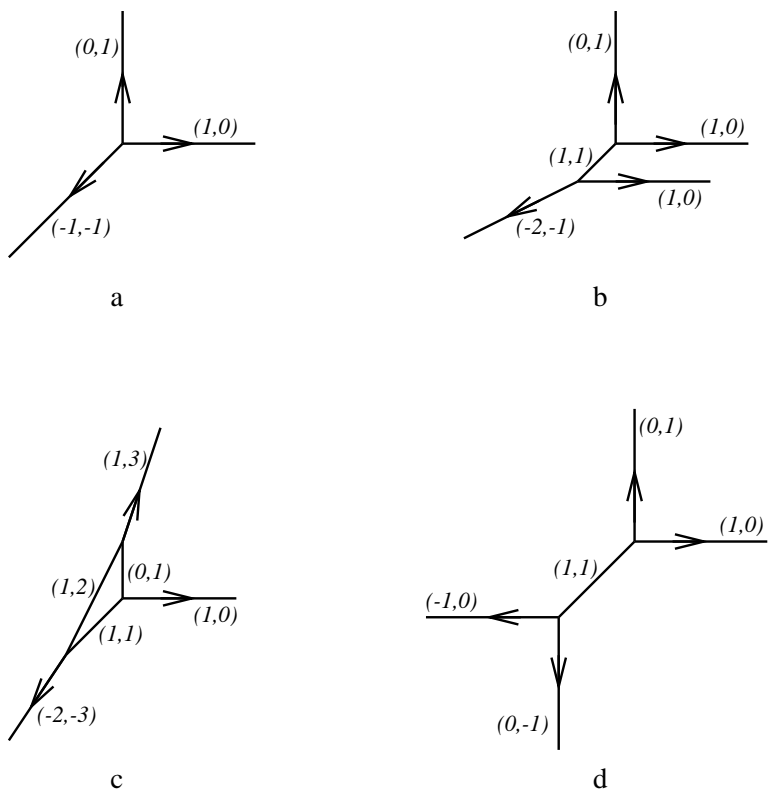}}
\caption{Examples of simple string webs: (a) a 3-string
junction, (b) a web with an internal string, (c) a web with
an internal face,
(d) another web with an internal string.}
\end{figure}

In the lift to M-theory string webs correspond to
membranes wrapping
holomorphic curves. The number of external strings is identified with the number
of boundaries $E_{ext}=b$, and the number of internal faces is identified with
the genus of the curve $F_{int}=g$.

\subsection{grid diagrams}

\noindent
A grid diagram
(similar diagrams appear in toric geometry)
is defined on a 2d square integer lattice. We shall denote its components by
{\em points, lines and polygons}. The diagram consists of points, which lie on
the grid, and of lines joining them. The contour of the diagram is convex, and
so are its internal polygons. There may be more conditions on the diagram, but
rather than state all of them, we shall describe how to build it.

To each string web one can associate a {\em dual} grid diagram, by exchanging
faces with points, $(p,q)$ edges (strings) with $\pm(-q,p)$ lines, and
vertices (junctions) with polygons. Consequently, the grid diagram consists of
$V$ adjacent polygons, and has $F_{ext}$ edge points, $F_{int}$ internal points,
$E_{ext}$ external lines, and $E_{int}$ internal lines. The grid diagrams dual
to the four string webs of figure~1 are shown in figure~2.
\begin{figure}[htb]
\epsfxsize= 4in
\centerline{\epsffile{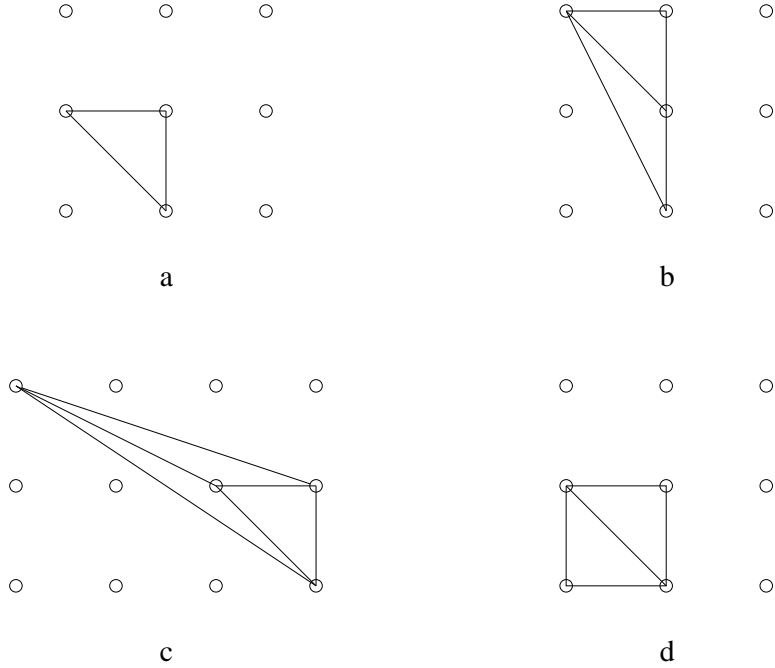}}
\caption{Grid diagrams dual to the webs in figure~1.
Note that (b) is a {\it quadrangle}.}
\end{figure}

To construct a grid diagram one starts by marking an
arbitrary point on the grid
which is chosen to correspond to some face in the web.
Crossing to an adjacent
face requires passing through a $(p,q)$ string, and thus
an orthogonal line
represented by the grid vector $\pm (-q,p)$ should be
marked, ending at a point
that represents the adjacent face. Consistency requires
that if we go around a
closed loop of faces in the web, circling a junction,
we will return to the same
point in the diagram. This is guaranteed by the charge
conservation property of
the junction (\ref{chargesum}). Convexity can be derived by
going around infinity in
the web.

\subsection{degenerations}

\noindent It may be the case that some of
the internal edges of a web have a
vanishing length. We shall refer to this as
a {\it degeneration} of the web.
In some cases degenerations may have interesting
physical consequences, for example:
\begin{itemize}
 \item External (figure~1b) or internal edges may overlap.
  In this case a web with $E_{ext}$ external
  ($E_{int}$ internal) edges degenerates into what
  looks like a web with less than $E_{ext}$ external
  ($E_{int}$ internal) edges, where
  some of them have non-co-prime $(p,q)$ charges.
  In the dual grid diagram such a potential degeneration
  corresponds to a point in the interior of
  an external (internal) line, {\it e.g.} figure~2b.
  For 5-brane webs, overlapping external and internal
  edges correspond to enhanced global
  and gauge symmetries, respectively, in the five
  dimensional field theories \cite{AH,AHK}.
  For string webs, overlapping external edges
  will be relevant when we
  discuss 1/4 BPS states in $N=4$ $SU(N_c)$ SYM, since
  they can end
  on the same D3-brane, and thus give rise to non-co-prime
  world-volume charges.
\item An internal face may shrink, {\it e.g.} figure~1c.
  For 5-brane webs this corresponds to a
  fixed point
  of the renormalization group in the five-dimensional
  field theory.
\item A web may become reducible, {\it e.g.} figure~1d,
  and therefore only marginally bound.
For 5-brane webs this would be a flop transition in the corresponding
Calabi-Yau.

\end{itemize}
If the external strings of a string web are not
infinite, but rather terminate on D-branes
(specifically D3-branes), there will be another kind of
degeneration, corresponding to shrinking
{\it external} strings. At such a
degeneration, the string web becomes marginal,
and in fact decays into smaller webs.
This will be explained in section~3.

\subsection {bosonic zero modes}
\noindent
A bosonic zero mode (BZM) is an infinitesimal deformation of the web which does
not change its mass
(see \cite{CallanThor} for a related study of the string web worldsheet).
Equivalently it is a solution of the scalar Laplace
equation on the web. Solutions on the entire web can be constructed by gluing
together the solutions for the individual edges (strings) at the vertices
(junctions). The latter are given by a constant 8-vector $X$ on each string (we
do not need to consider rotations here). These decompose into 7-vectors normal
to the plane of the web, $X_N$, and scalars tangent to the plane of the web (and
normal to the string) $X_T$,
\be
X \to X_N + X_T \; .
\ee
These variables are constrained at the junctions by
\bea
X_N^1=X_N^2=X_N^3 \label{BZM_vertexN} \\
\sum_{i=1}^{3}(\pm){\left|p_i + \tau q_i \right|X_T^i=0}\; ,
\label{BZM_junctionT}
\eea
where the superscript $i=1,2,3$ labels the strings that
meet at the junction. The
sign is determined by whether the planar deformation of a
given string is
clockwise or counterclockwise with respect to the junction.

The condition (\ref{BZM_vertexN}) implies that the web has a total of 7 normal
BZM. These correspond
to the motion of the entire web in the transverse
directions. The condition (\ref{BZM_junctionT}) gives $V$ (independent) junction
constraints for $E$ variables. Thus the number of planar BZM is
$E-V=F-1=F_{int}+E_{ext}-1$, where we have used (\ref{euler}) and
(\ref{externals}).

We distinguish between {\it local BZM} and {\it global BZM}. Define the {\it
support} of a zero mode to be the union of all strings on which the mode does
not vanish. A local BZM does not move the external strings, so these are not
included in its support. The number of global BZM is $E_{ext}-1$, which includes
the two translations in the plane of the web. From the counting we see that each
internal face contributes one local BZM, which is realized as the
shrinking/expansion mode of that face. In general, the web might have ``hidden
faces", and one should instead count the internal points in the dual grid
diagram.

By supersymmetry, string webs
possess fermionic zero modes (FZM) as well. Furthermore,
if the external strings
terminate on D3-branes, the number of both BZM and
FZM will change due
to the boundary conditions. This will be shown in section~3.

\subsection{classification of string webs}

\noindent As we have mentioned, the simple 3-string junction
(figure~1a) is one component of an $SL(2,\bbbz)$ multiplet of 3-string
junctions. Likewise all string webs belong to $SL(2,\bbbz)$
multiplets. These
can be classified using $SL(2,\bbbz)$ invariants of string webs. The obvious
invariants are the web variables $V, E_{ext,int}$, and $F_{ext,int}$. These are
however not independent, as they are related by (\ref{euler}),(\ref{externals}),
and (\ref{3junction}). The two independent invariants can be taken to be
$E_{ext}$ and $F_{int}$.

Consider for example string webs with $E_{ext}=3$.
These are
generalizations of the 3-string junction, to include an
arbitrary number of internal
faces $F_{int}$. Label the external strings
$(p_1,q_1), (p_2,q_2)$, and $(p_3,q_3)$.
By charge conservation $p_3=-p_1-p_2$ and $q_3=-q_1-q_2$.
One can define an $SL(2,\bbbz)$
invariant
intersection number \cite{gzunpublished}
\be
  {\cal I} = \left|
  \begin{array}{ll}
   p_1 & p_2 \\
   q_1 & q_2
  \end{array}
  \right| = p_1q_2 - p_2q_1\; .
\ee
This is simply twice the (oriented) area of the
triangle in the corresponding grid
diagram, and is related to the number of internal points
$F_{int}$ as
follows \footnote{For an $E_{ext}$ sided polygon with
$F_{int}$ internal points,
the area is given by $A=F_{int} + E_{ext}/2 - 1$.}
\be
 {\cal I} = 2A = 2(F_{int}+1/2) \; .
\ee
We shall refer to this as an ``area'' type invariant.
For webs with more
external strings there will be several ``area'' invariants,
given by the areas of the triangular sub-polygons, modulo
constraints. For
example, webs with $E_{ext}=4$ have dual grid diagrams
with four external lines, {\it
i.e.} quadrangles \footnote{These may be degenerate,
as in figure~2b.}.
These can be divided into two ``triangles'' in two ways,
corresponding to the two diagonals
\footnote{If the diagonals go through internal points,
these ``triangles'' may actually be higher polygons.}. 
Thus there are four
``triangular'' sub-polygons with areas
$A_1,\ldots,A_4$.
There is one
constraint coming from the one quadrangle,
namely $A_1+A_2 = A_3+A_4$.
That leaves
three invariants, which can be
related to the three independent intersection numbers
\be
 {\cal I}_{12} = \left|
   \begin{array}{ll}
     p_1 & p_2 \\
     q_1 & q_2
   \end{array}\right|
 \qquad
 {\cal I}_{23} = \left|
   \begin{array}{ll}
     p_2 & p_3 \\
     q_2 & q_3
   \end{array}\right|
 \qquad
 {\cal I}_{31} = \left|
   \begin{array}{ll}
     p_3 & p_1 \\
     q_3 & q_1
   \end{array}\right| \; .
\ee
This generalizes to arbitrary values of $E_{ext}$.
The number of independent
intersection numbers is given by
\be
 \left( \begin{array}{c}
         E_{ext}-1 \\ 2
        \end{array} \right) \; .
\label{intersections}
\ee
In the grid diagram this is understood as follows.
One counts the number of triangular sub-polygons
and subtracts the number of constraints. These
are given by the number of quadrangular sub-polygons,
minus constraints on constraints, etc.
The result is
\be
 \left( \begin{array}{c}
         E_{ext} \\ 3
        \end{array} \right)
 - \left( \begin{array}{c}
         E_{ext} \\ 4
        \end{array} \right)
 +  \left( \begin{array}{c}
         E_{ext} \\ 5
        \end{array} \right)
 - \cdots
 = \sum_{k=3}^{E_{ext}} (-1)^{k+1}
    \left( \begin{array}{c}
         E_{ext} \\ k
        \end{array} \right) \; .
\label{areas}
\ee
One can then check using the binomial formula that
this is indeed equal to (\ref{intersections}).

There are additional invariants which are not related to
areas of polygons (or
intersection numbers). For the
case $E_{ext}=3$ there is one such invariant, which is
understood as follows \cite{gzunpublished}.
Using $SL(2,\bbbz)$ we can transform
one of the external
strings into a $(1,0)$ string. The charges of the external
strings are then given by
\be
 (1,0)\; , \qquad (p,q)\; , \qquad (-p-1,-q) \; .
\ee
The integer $q$ is equal to the ``area'' invariant ${\cal I}$,
and its value is arbitrary. The other invariant is
equal to the integer $p$, whose value determines the
shape of the grid diagram (see figure~3). Hence it is
referred to as a ``shape'' invariant.
There are residual $SL(2,\bbbz)$ transformations
which preserve the $(1,0)$ string. These are given by
\be
 \left(
 \begin{array}{cc}
  1 & n \\
  0 & 1
 \end{array}\right) \; , \qquad n\in\bbbz \; .
\ee
Under such a transformation $p\rightarrow p+nq$, so only values of $p$ in
$[0,q-1]$ are $SL(2,\bbbz)$ inequivalent. The allowed values of $p$ are actually
more restricted, since both $(p,q)$ and $(-p-1,-q)$ must be co-prime. For $q>0$
and prime, the only excluded values are $p=0$ and $p=q-1$. For $q$ non-prime
there will be additional exclusions. In the grid diagram the exclusions can be
understood as the appearance of additional edge points as the shape of the
diagram is changed (figure~3).
\begin{figure}[htb]
\epsfxsize= 4.5cm
\centerline{\epsffile{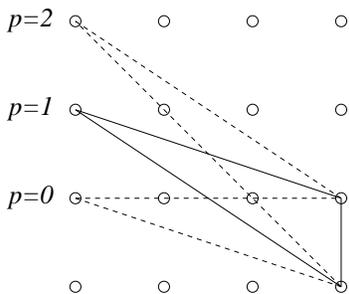}}
\caption{Grid diagrams with $E_{ext}=3, q=3$ and varying shapes.
The cases
$p=0,2$ are excluded as they contain new edge points, and
therefore really have $E_{ext}>3$.}
\end{figure}

\section{1/4 BPS States in $SU(N_c)$ SYM}
\setcounter{equation}{0}

\noindent In this section we will generalize
the results of \cite{Bergman} to 1/4 BPS states of
$SU(N_c)$ SYM for arbitrary $N_c$, carrying
arbitrary electric and magnetic charges, and with an
arbitrary value of
$\theta_{YM}$. States in $SU(N_c)$ SYM are characterized
in part by their
$(N_c-1)$-dimensional electric and magnetic charge
vectors $(\bp,\bq)$. These
charges are related via the vacuum expectation value
of the sextuplet Higgs
field in the Cartan subalgebra 
$\langle\vec{\bPhi}\rangle = \vec{\bv}$ to
six-dimensional electric and
magnetic charge vectors, given by
\footnote{We shall use {\bf boldface} to
denote $(N_c-1)$-dimensional vectors, and an
$\vec{a}rrow$ to denote
six-dimensional vectors.}
\be
 \vec{Q}_E = g_{YM}\sum_{a=1}^{N_c-1}\left(
     p^a\bbeta^{(a)}\cdot\vec{\bv}
     + {\theta_{YM}\over 2\pi}
    q^a\bbeta^{(a)*}\cdot\vec{\bv}\right)
  \quad , \quad
 \vec{Q}_M = {4\pi\over g_{YM}}
  \sum_{a=1}^{N_c-1} q^a\bbeta^{(a)*}\cdot\vec{\bv} \; ,
\ee
where $\bbeta^{(a)}$ and $\bbeta^{(a)*}$
($a=1,\ldots,N_c-1$) are the
simple roots and simple co-roots,
respectively, of $SU(N_c)$.
Since $SU(N_c)$ is simply-laced $\bbeta^{(a)}
=\bbeta^{(a)*}$.
The central charges of the $N=4$ superalgebra are
given by \cite{osborn}
\be
 Z_{\pm}^2 = |\vec{Q}_E|^2 + |\vec{Q}_M|^2
       \pm 2|\vec{Q}_E\times \vec{Q}_M| \; ,
\label{Zpm}
\ee
and therefore the BPS bound on the mass of charged states
is
\be
 M_{(\bp,\bq)} \geq Z_+ \; .
\ee
There are two possible kinds of BPS states.
The $W$-bosons, monopoles,
and usual dyons have $\bp\propto\bq$, and therefore
$\vec{Q}_E\propto\vec{Q}_M$. These states preserve one
half of the underlying supersymmetry, {\it i.e.} eight
supercharges, and transform in a short ($2^4$)
representation of the superalgebra,
with $j_{max}=1$ (vector multiplets).
Hence they are referred
to as 1/2 BPS states.
On the other hand, if $N_c>2$, it is possible to have
$\bp\propto\!\!\!\!\!\!/ \,\,\,\bq$, and therefore
$\vec{Q}_E\propto\!\!\!\!\!\!/ \,\,\,\vec{Q}_M$.
Such states would preserve only one quarter of the
supersymmetry, {\it i.e.} four supercharges, and therefore
transform in a medium ($2^6$) representation of the
superalgebra, with $j_{max}=3/2$.
We refer to these as 1/4 BPS states.

\subsection{geometry and BPS mass}

\noindent In the D-brane construction we have $N_c$
parallel D3-branes,
with positions
$\vec{R}_a$ in the transverse $\bbbr^6$ space.
This gives a world-volume gauge group
$U(N_c)\sim SU(N_c)\times U(1)$. The $U(1)$
factor corresponds to the c.o.m. degrees of
freedom which we can ignore.
Let us fix the position of one of the D3-branes at the
origin $\vec{R}_{N_c}=0$, and parameterize the moduli
space by the positions of the $N_c-1$ others. This
also means that we will parameterize the Cartan
subalgebra of $SU(N_c)$ according to the $N_c-1$ free
D3-branes.

The SYM coupling constant and theta angle are related to
the string coupling
constant and RR scalar as
\be
 \tau = {4\pi i\over g_{YM}^2} + {\theta_{YM}\over 2\pi} =
        {i\over g_s} + a \; .
\ee
The positions of the D3-branes are related to VEV's
of the Higgs field as follows
\be
 \vec{R}_a = 4\pi^{3/2}\alpha' \bbeta^{(a)}\cdot
   \vec{\bv} \; .
\ee
A simple way to determine the precise numerical factor is
to compare the masses
of the $W$-boson and monopole to the masses of a $(1,0)$
string and a $(0,1)$
string between two D3-branes. We shall henceforth set
$2\pi\alpha'=1$, and
assume that $g_s=1$ and $a=0$.
The generalization to other values is
achieved by simply replacing
\be
  \bp \longrightarrow g_s(\bp + a\bq) \; .
\ee
We can now express the central charge $Z_+$ in terms of
the D3-brane positions
$\vec{R}_a$:
\be
 Z_+^2 =   \sum_{a,b=1}^{N_c-1}
  (p^ap^b + q^aq^b)\vec{R}_a\cdot\vec{R}_b
  + 2\left|\sum_{a,b=1}^{N_c-1}p^aq^b
   \vec{R}_a\times\vec{R}_b \right| \; .
\label{Zgeo}
\ee
We shall assume that $\bp$ and $\bq$ have no
common vanishing components. If they did, the problem
would reduce to finding BPS states in
$SU(N_c-k)$, where $k$ is the number of common vanishing
components.
We are therefore interested in configurations where
strings end on {\it all} $N_c$ D3-branes.

Let us first consider degenerate string webs, in
which all internal strings have vanishing length,
and therefore all the external strings meet at a
point (figure~4a).
We will then generalize to include non-vanishing
internal strings (figure~4b), as well as non-vanishing
internal faces (figure~4c).
The external strings carry charges
$(p_1,q_1),\ldots,(p_{N_c},q_{N_c})$, and
lie along the vectors
$\vec{A}_a$ ($a=1,\ldots,N_c$). The geometrical
relations that follow from figure~4a are given by
\be
 |\vec{R}_a - \vec{R}_b|^2
  = |\vec{A}_a - \vec{A}_b|^2
  = A_a^2 + A_b^2 -2A_aA_b\cos\theta_{ab}
  \qquad (a,b=1,\ldots,N_c)\;,
\label{geometry1}
\ee
where $\theta_a=\mbox{Arg}(\vec{A}_a)$, and
$\theta_{ab}\equiv\theta_a - \theta_b$.
The supersymmetry condition (\ref{angles}) implies that
(with $g_s=1$ and
$a=0$)
\be
 \tan\theta_a = {q_a\over p_a} \; ,
\label{anglecondition}
\ee
and therefore that
\be
 \cos(\theta_a-\theta_b) = {p_ap_b + q_aq_b\over
   \sqrt{(p_a^2+q_a^2)(p_b^2+q_b^2)}}
    \qquad , \qquad
 \sin(\theta_a-\theta_b) = {p_bq_a - p_aq_b\over
   \sqrt{(p_a^2+q_a^2)(p_b^2+q_b^2)}} \; .
\label{geometry2}
\ee
Using (\ref{geometry1}) and (\ref{geometry2}) in
(\ref{Zgeo}) we find
\be
 Z_+ =  \sum_{a=1}^{N_c}
    A_a\sqrt{(p_a^2 + q_a^2)} \; ,
\label{agreement}
\ee
where $p_{N_c}=-\sum_{a=1}^{N_c-1}p_a$, and similarly for
$q_{N_c}$.
This is precisely the mass of the string web in figure~4a.

For string webs that have internal strings, but no
(finite size) internal faces, a generic geometry is shown in
figure~4b. The internal strings carry charges
$(r_1,s_1),\ldots,
  (r_{E_{int}},s_{E_{int}})$, and
lie along
the vectors $\vec{B}_\alpha$ ($\alpha=1,\ldots,E_{int}$).
Generically, a path from one D3-brane to another
goes through $k$ internal strings
$\vec{B}_{\alpha_1},\ldots,\vec{B}_{\alpha_k}$,
so the geometrical relations are now
\be
  |\vec{R}_a - \vec{R}_b|^2 =
   |\vec{A}_a - \vec{A}_b + \sum_{i=1}^k
    (\pm)\vec{B}_{\alpha_i}|^2  \; ,
\label{geometry3}
\ee
where the sign depends on the orientation
of the vector $\vec{B}_{\alpha_i}$ relative to
the path.
The supersymmetry condition for the internal
strings is
\be
 \tan\phi_\alpha = {s_\alpha\over r_\alpha} \; ,
\ee
where $\phi_\alpha = \mbox{Arg}(\vec{B}_\alpha)$,
and there are therefore relations analogous
to (\ref{geometry2}) for
$(\phi_\alpha - \phi_\beta)$
and $(\theta_a - \phi_\beta)$.
These, together with (\ref{geometry3}), give
\be
 Z_+ =   \sum_{a=1}^{N_c} A_a \sqrt{p_a^2 + q_a^2} +
   \sum_{\alpha=1}^{E_{int}}
  B_\alpha \sqrt{r_\alpha^2 + s_\alpha^2}
  \; ,
\label{result}
\ee
which is precisely the mass of the string web.

Finally, in the most general case string webs can
have internal faces as well (figure~4c). The geometrical
relations are still (\ref{geometry3}), but the steps
to (\ref{result}) are a little more complicated,
as some internal strings may not appear in any paths.
On the other hand, we already know that shrinking an
internal face of a string web corresponds to a bosonic
zero mode (we will see that this persists in the presence
of D3-branes as well in subsection~3.3), so the mass
of the web is independent of this deformation.
Since shrinking the internal faces to zero size reduces
the web to the previous case, the agreement holds
in the presence of finite internal faces as well,
and therefore for generic string webs.
\begin{figure}[htb]
\epsfxsize=5.5in
\centerline{\epsffile{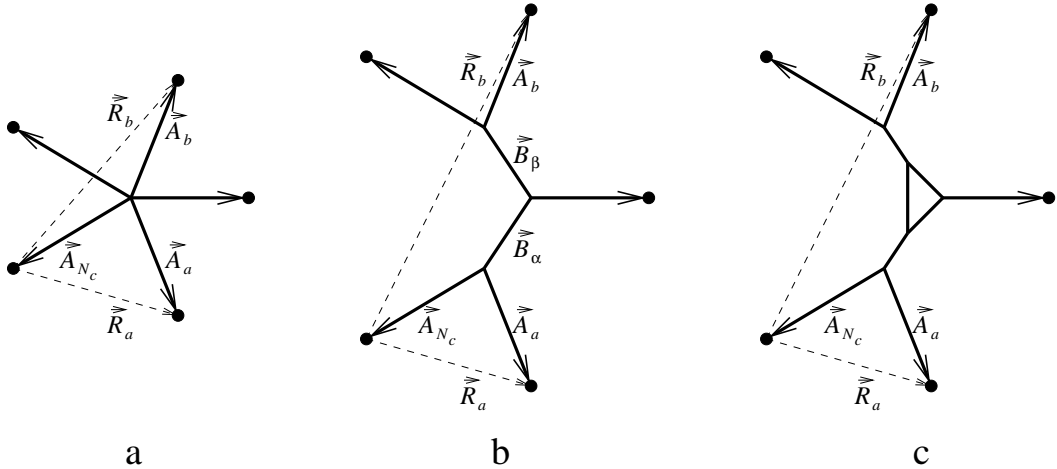}}
\caption{1/4 BPS states from string webs: (a)
completely degenerate web, (b) web without internal faces,
(c) generic web.}
\end{figure}

Note that the agreement holds regardless of whether
the charges $p_a$ and $q_a$ are co-prime or not,
{\it i.e.} whether
some external strings are overlapping.
We expect a 1/4 BPS state in the field theory
(modulo marginal stability, soon
to be addressed) for all sets $\{(p_a,q_a)\}$ of
charges satisfying $\sum^{N_c} p_a =
\sum^{N_c} q_a = 0$, as long as the charges of the different
external strings do not
have a common divisor. If there is a common divisor,
say $k$, the configuration
is reducible to $k$ irreducible ones, corresponding to
$k$ 1/4 BPS states,
and the state could be at most marginally bound.

An alternative derivation for the equivalence of the
BPS formula and the mass of
the web is given in Appendix~A.

\subsection{examples}

\noindent Let us consider a few examples of
1/4 BPS states in $SU(3)$ and $SU(4)$ SYM.
We shall use the four examples of string webs
from the previous section (figure~1),
and assume that the
external strings terminate on D3-branes.
\begin{description}
\item{\it Example~1}
The 3-string junction in figure~1a corresponds to the state
$((p_1,q_1),(p_2,q_2))=((1,0),(0,1))$ in $SU(3)$, which was studied in
\cite{Bergman}.
\item{\it Example~2}
The web in figure~1b has four external
strings, $(0,1)$, $(-2,-1)$, and two $(1,0)$'s.
When these end on four D3-branes a 1/4 BPS state
with charges $((0,1),(1,0),(1,0))$ results in the
world-volume $SU(4)$ theory.
In the degenerate case, {\it i.e.} when the internal
$(1,1)$ string shrinks, the two $(1,0)$ strings
coincide, and can therefore terminate on a single
D3-brane. This gives rise to a 1/4 BPS state in
$SU(3)$, with charges $((0,1),(2,0))$.
\item{\it Example~3}
The web in figure~1c has three external strings and a single internal face. It
gives rise to a 1/4 BPS state in $SU(3)$ SYM with charges $((1,0),(1,3))$. This
state has a single bosonic zero mode, which must be quantized.
\item{\it Example~4}
The web in figure~1d has four external strings, and therefore gives rise to a
1/4 BPS state in $SU(4)$ SYM. The charges of this state are
$((1,0),(0,1),(0,-1))$.
\end{description}
All of the above states preserve 1/4 of the underlying supersymmetry of $N=4$
SYM, and therefore would seem to transform in medium sized ($2^6$) multiplets,
with $j_{max}=3/2$. However, as we shall soon see, there are generically
additional fermionic zero modes beyond those corresponding to the 12 broken
supersymmetries. These generate additional degeneracy, and thus increase the
size of the multiplets, and with it $j_{max}$.

\subsection{fermionic zero modes}
\label{FZM}

In section~2 we analyzed the bosonic zero modes (BZM) of
string webs. Here we
shall first extend the discussion of the BZM to include
D3-brane boundaries, and
then we shall find the fermionic zero modes (FZM).

Adding the D3-branes further decomposes $X_N$, the deformation 7-vector normal
to the web, into a 4-vector perpendicular to the D3-branes, $X_{N\perp}$, and a
3-vector parallel to the branes, $X_{N\parallel}$
\be
X_N \to X_{N\perp} + X_{N\parallel}\; .
\ee
The boundaries constrain the strings ending on the D3-branes, {\it i.e.} the
external strings, to move only in worldvolume directions, so
\be
 X_T\Big|_{ext} = X_{N\perp}\Big|_{ext} = 0 \; .
\label{BZM_boundary}
\ee
This eliminates the planar global BZM, leaving only the $F_{int}$ local BZM. It
also eliminates 4 of the 7 normal BZM, leaving 3 which account
for translations along the D3-branes.

The local BZM actually need to be complexified to give the full smooth moduli
space of the web. Though we will not use this fact, note that in a web
description there is a singularity (or boundary) in the moduli space when an
internal face expands until it hits a D3-brane. Experience with geometrical
descriptions of the system \cite{Witten_phase,BK5vM,LV} teaches us that each
modulus in the web picture could be paired with a complex phase.

Let us now derive the equations for the fermionic zero modes (FZM). As for the
BZM, we can solve the Dirac equation on the web by gluing together the solutions
on individual strings. The latter are constant 16 component spinors $\Psi$,
which transform in the same representations as the broken supersymmetries of the
respective strings. We would like to describe the constraints on these variables
at the junctions and at the boundaries. The spinors decompose under the
transverse $SO(7)$ subgroup as
\be
\Psi \to \Psi_T + \Psi_N\; ,
\ee
where $\Psi_T,\Psi_N$ are the superpartners of $X_T,X_N$,
respectively, under
the 8 supersymmetries preserved by the web
(the number of bosons and fermions in
a multiplet does not have to match as the
zero modes are quantum mechanical,
rather than field theoretic, variables.).

In the presence of D3-branes only 4 supersymmetries are unbroken. However, since
junctions still preserve the original 8 supersymmetries {\it locally}, one can
derive the fermionic junction conditions by acting on the bosonic ones
(\ref{BZM_vertexN}), (\ref{BZM_junctionT}) with these supersymmetries:
\bea
\Psi_N^1=\Psi_N^2=\Psi_N^3 \label{FZM_vertexN} \\
\sum_{i=1}^{3}(\pm){\left|p_i + \tau q_i \right| \Psi_T^i=0}\;.
\label{FZM_vertexT}
\eea
We use the same sign conventions as in the bosonic case.

After introducing the D3-branes
the spinors are further decomposed under $SO(7)
\to SO(3)
\times SO(4)$ as
\bea
\Psi_T &\to & ({\bf 2},({\bf 2},{\bf 1}))
   +({\bf 2},({\bf 1},{\bf 2}))
=\Psi_{T\perp}+\Psi_{T\parallel}
\nonumber \\
\Psi_N &\to & ({\bf 2},({\bf 2},{\bf 1}))
  +({\bf 2},({\bf 1},{\bf 2}))
=\Psi_{N\parallel}+\Psi_{N\perp}\;,
\eea
where $\Psi_{T\perp}$, $\Psi_{N\perp}$, and $\Psi_{N\parallel}$
are the
superpartners of $X_T$, $X_{N\perp}$, and $X_{N\parallel}$,
respectively, under the 4 globally
preserved supersymmetries
($\Psi_{T\parallel}$ has no superpartner).
At a given boundary there
are 8 locally preserved supersymmetries
(supersymmetries that are preserved both by the D3-brane and
the string ending on it).
By acting with these supersymmetries on the bosonic boundary constraint
(\ref{BZM_boundary}) we get the fermionic boundary constraint
\footnote{Actually it is sufficient in this case to act
only with the 4 globally preserved supersymmetries.}:
\be
\Psi_{T\perp}\Big|_{ext}=\Psi_{N\perp}\Big|_{ext}=0 \; .
\label{FZMboundary}
\ee

Having derived the equations for the FZM we would like to
describe their
solutions. First we shall count the solutions directly from
the equations, then
we shall confirm the result by generating FZM from BZM by
acting with supersymmetries.
Equations (\ref{FZM_vertexN}) and (\ref{FZMboundary}) imply that
$\Psi_{N\perp}=0$ on all the strings. On the other hand, the solution
$\Psi_{N\parallel}=const$ (same constant on all the strings)
describes 4
fermionic zero modes. For $\Psi_{T\parallel}$ there are $4E$
variables and $4V$
(independent) constraints from (\ref{FZM_vertexT}),
which gives $4(E-V) =
4(F_{int}+E_{ext}-1)$ FZM. For $\Psi_{T\perp}$, on the other hand,
there are
only $4E_{int}$ variables, since the ones on the external strings must vanish
(\ref{FZMboundary}). Eq.~(\ref{FZM_vertexT}) gives $4V$ constraints on
$\Psi_{T\perp}$, but they are not independent. The constraints are related by a
single (spinor) equation. This can be seen by summing the constraints in
(\ref{FZM_vertexT}) over all the vertices, and noting that the contributions of
internal strings drop out since they always contribute to two vertices with
opposite sign, and that the contribution of the external strings vanishes due to
(\ref{FZMboundary}). There are therefore $4(V-1)$ independent constraints on the
$4E_{int}$ variables, leaving $4(E_{int} - V + 1) = 4F_{int}$ FZM. Summing all
the contributions gives a total of $8F_{int}+4E_{ext}$.

The FZM can also be obtained directly from the BZM using certain supersymmetry
(not necessarily unbroken) transformations. This method has the advantage of
giving the metric on moduli space. Consider first a local BZM. The support of
such a mode does not include external strings, which end on D3-branes, and
therefore it preserves 8 supersymmetries. By applying these to the BZM we get
constant fermionic variables that
satisfy the junction constraint
(\ref{FZM_vertexT}) by construction.
Since there are $F_{int}$ local BZM, the above procedure
produces $8F_{int}$ FZM
\footnote{Another way to see that each local BZM gives
8 FZM is to consider a 5-brane \pq web, where each face is known to give rise to
a vector multiplet \cite{AHK}. A vector multiplet in 5d has one scalar, which is
the BZM, 4 fermionic degrees of freedom coming from 8 fermionic variables, which
are the 8 FZM, and 3 vector degrees of freedom which do not appear in the
quantum mechanics.}.
Similarly, one can obtain the 4 FZM corresponding
to $\Psi_{N\parallel}=const$ by acting on the
translational BZM $X_{N\parallel}=const$ with the 4 globally
preserved supersymmetries.

The remaining FZM can be generated from the $E_{ext}-1$ planar global BZM of the
unbounded web (section~2.4), even though they no longer exist in the presence of
D3-branes. Let us act on these BZM, {\it i.e.} $X_T$ with $X_T|_{ext}\neq
0$, with the 4 supersymmetries which are preserved by the web, but {\it broken}
by the D3-branes. This produces constant fermionic variables which satisfy the
junction conditions. Furthermore, as these supersymmetries transform in the
$({\bf 2},({\bf 1},{\bf 2}))$ of $SO(3) \times SO(4)$, we obtain
$\Psi_{T\parallel}$ rather than $\Psi_{T\perp}$. The boundary condition
$\Psi_{T\perp}|_{ext}=0$ can therefore be satisfied despite the fact that
$X_T|_{ext}\neq 0$. Since both the junction and boundary conditions are
satisfied, these are indeed FZM, and there are $4(E_{ext}-1)$ of them. The total
number of FZM is then
\be
n_{FZM}=8F_{int} + 4 + 4(E_{ext}-1) = 8F_{int}+4E_{ext} \; ,
\label{n_FZM}
\ee
in agreement with the previous count.

An advantage of this method of generating FZM is that the fermionic mass matrix
(the metric on the fermionic coordinates of moduli space) is automatically
determined. The mass of an FZM is given by the mass of the BZM that was used to
generate it. For BZM, the metric on moduli space $g_{ij}$ can be read off from
the expression for the kinetic energy
\be
g_{ij}=\sum_{a \in strings}{m_a X_{ai} X_{aj} }\; ,
\ee
where $i,j$ label the BZM. The fermionic kinetic term is analogously
$g_{ij}=\sum {m_a \bar{\Psi}_{ai} \Psi_{aj} }$. We see here
explicitly that FZM generated from BZM will have the same metric. This applies
not only to FZM generated by locally unbroken supersymmetries, but also to the
FZM that are generated by broken supersymmetries.

Let us illustrate the result in (\ref{n_FZM}) by some examples:
\begin{itemize}
\item{The $W$ boson is just an open string, and
therefore has $E_{ext}=2, F_{int}=0$. Eq.~(\ref{n_FZM}) gives 8 FZM, which is
consistent with it being $1/2$ BPS and preserving $8$ supersymmetries. The FZM
generate a $2^4$ (short) representation of the superalgebra with $j \leq 1$.}
\item{The 3-string junction
(figure~1a) has $E_{ext}=3, F_{int}=0$, and therefore
12 FZM, which agrees with the
number of broken supersymmetries. These generate a $2^6$
(medium) representation with
$j\leq 3/2$.}
\item{The states arising from the string webs in
figures~1b,d have $E_{ext}=4, F_{int}=0$, and therefore 16 FZM, 4 of which do
not correspond to broken supersymmetries. These generate a $2^8$ representation
with $j \leq 2$.}
\item{The state in figure~1c has $E_{ext}=3, F_{int}=1$, and
therefore 20 FZM. The 4 modes related (by supersymmetry) to translations along
the D3-branes have masses which are independent of the bosonic moduli, while the
masses of the other modes depend on the size of the internal face, as can be
seen for the corresponding BZM. The former generate a $2^2$ representation,
which is tensored with the ground states of the quantum mechanics on a 1 complex
dimensional manifold with the remaining 16 FZM.}
\end{itemize}

One can also use the above methods to count fermionic zero modes of string webs
with no D3-brane boundaries.
One gets
\be
\textnormal{ web without boundaries: } n_{FZM}=8F_{int}+8+8(E_{ext}-1)
=8F_{int}+8E_{ext},
\ee
where $8F_{int}$ are obtained from local BZM, 8 are obtained from translations
in the 7 transverse directions, and $8(E_{ext}-1)$ are obtained from planar
global BZM.

For periodic string webs \cite{Sen_net} (a net on a torus), the BZM consist of 7
translations transverse to the web, and $F+1$ modes in the plane of the web, two
of which are translations, where
$F$ is the number of faces in a unit cell. Varying the
translations with respect to the 8 supersymmetries preserved by the web gives 8
FZM, and varying the planar modes gives $8(F+1)$ FZM. The former are the
solutions of the $\Psi_N$ equations (\ref{FZM_vertexN}), and the latter solve
the $\Psi_T$ equations (\ref{FZM_vertexT}). Summing the FZM gives
\be
\textnormal{web on torus: } n_{FZM}=8(F+2).
\ee

In the lift to M-theory the string web becomes a smooth membrane, with genus
$g=F_{int}$ and $b=E_{ext}$ boundaries. Fermionic zero modes correspond to zero
eigenvalues of the Dirac operator on the membrane. For a web with D3-brane
boundaries we have $n_{FZM}=4(2-\chi)$, where $\chi$ is the Euler character of
the membrane. For a periodic web the genus is $g=F+1$, and we have
$n_{FZM}=4(4-\chi)$

\subsection{classification of BPS states}

\noindent The BPS states of $N=4$ $SU(N_c)$ SYM
transform in irreducible
representations of the $SL(2,\bbbz)$ duality group.
They can therefore be classified using $SL(2,\bbbz)$
invariants. The simplest invariant is the number
of {\it independent} non-vanishing $(p,q)$ charges.

The gluons in the Cartan subalgebra carry no charges
and are therefore invariant under $SL(2,\bbbz)$. Since they
are massless, they are 1/2 BPS states.

States carrying a single independent charge are
also 1/2 BPS states, but are generically massive.
Note that there
could still be two charges whose sum vanishes, as for the
$((1,0),(-1,0))$
$W$-boson of $SU(3)$. All of these 1/2 BPS states can be
obtained from the
perturbative ones, namely from the $N_c(N_c-1)$
$W$-bosons,
by the action of the
duality group. This will include all the monopoles and
dyons carrying parallel
and co-prime charges $(\bp,\bq)$.

The 1/4 BPS states are not related to the perturbative
states by any
$SL(2,\bbbz)$ element, and therefore lie on separate
$SL(2,\bbbz)$
orbits \cite{dyons}.
As these states correspond to string webs, their
classification will be similar, but not identical, to the
classification of
string webs. String webs with overlapping external strings
give rise to multiple
BPS states, depending on whether overlapping  external strings
end on the same
D3-brane or not. Non-degenerate string webs with $E_{ext}$
external strings will
give rise to 1/4 BPS states
carrying $E_{ext}-1$ independent $(p,q)$ charges, all of
which are co-prime. On the other hand, degenerate string webs
with overlapping
external strings will give rise to these, as well as to 1/4 BPS
states carrying
less charges, some of which are non-co-prime, as in example~2 of
section~3.2
(figure~1b). In counting 1/4 BPS states we must allow for
non-co-prime charges,
as long as there isn't a common divisor for all the charges.
The latter case
will correspond to a reducible string web, and therefore to a
multi-particle
state.

Consider for example 1/4 BPS states in $SU(3)$, or equivalently
states carrying two independent charges in any $SU(N_c)$.
Following the same steps as in section~2, we use
$SL(2,\bbbz)$ to transform one of the charges to
$(k,0)$. This leaves us with the second charge $(p,q)$.
The charge $q$ is arbitrary,
but $p$ is restricted to the interval
$[0,q-1]$ as before. Unlike in the
classification of string webs however, there are no
further restrictions on $p$. In particular, neither
$(p,q)$ nor $(-p-k,-q)$ are required to be co-prime.
In summary, the $SL(2,\bbbz)$ orbits of 1/4 BPS states
carrying two independent charges are classified by two
arbitrary integers $k,q$ and an integer $p\in [0,q-1]$.
It would be interesting to find a geometrical
interpretation for $k$ and $q$, as was
done for string webs in section~2.5.
One could attempt to derive invariants by referring to the
grid diagram, which for the
above states is a triangle \footnote{This includes degenerate
polygons which look like triangles, {\it e.g.}
figure~2b.}. The obvious
invariants are the
number of internal points, and the number of edge points on
each of the lines.
These are however not independent, and non-trivially
related to $k$ and $q$.

\subsection{curves of marginal stability and decay of BPS
states}

\noindent Consider a BPS state in $N=4$ $SU(N_c)$
SYM with charges $((p_1,q_1),\ldots,(p_{N_c-1},q_{N_c-1}))$.
Assume the charge vector $\bp$ is decomposed as
$\bp=\bp'+\bp''$, and similarly for $\bq$. The mass of the state in question
satisfies a triangle inequality
\be
M_{(\bp,\bq)} \le M_{(\bp',\bq')} + M_{(\bp'',\bq'')} \; .
\ee
There exists a subspace of the moduli space
where the above inequality
is saturated, and the BPS state is only marginally
stable against decay into the two
other BPS states.

This happens already for 1/2 BPS states if $N_c>2$.
Consider the $W$-bosons of
$SU(3)\rightarrow U(1)^2$. These carry charges
$\pm((1,0),(0,0)),
\pm((0,0),(1,0))$ and $\pm((1,0),(-1,0))$ under the
two $U(1)$'s,
and are represented by fundamental strings between
pairs of D3-branes.
Their masses are therefore given by
\begin{eqnarray}
 M_{\pm((1,0),(0,0))} &=& |\vec{R}_1| \nonumber \\
 M_{\pm((0,0),(1,0))} &=& |\vec{R}_2| \nonumber \\
 M_{\pm((1,0),(-1,0))} &=& |\vec{R}_1 - \vec{R}_2| \; .
\end{eqnarray}
Consider the subspace of moduli space defined by
$\vec{R}_2 = c\vec{R}_1$,
where $0<c<1$. This corresponds to choosing the Higgs
VEV's in the
two Cartan directions to be parallel.
In this subspace
\be
  M_{((1,0),(0,0))}  = M_{((0,0),(1,0))}  +
  M_{((1,0),(-1,0))} \; ,
\ee
so the $((1,0),(0,0))$ $W$-boson is marginally stable
against decay into the two
others. One can similarly find marginal stability subspaces
for the other two
$W$-bosons, and more generally for all 1/2 BPS states.
Since these subspaces always have a boundary however,
they do not divide moduli
space into disjoint regions, and therefore the 1/2 BPS
states are absolutely stable everywhere else.

The 1/4 BPS states exhibit marginal stability on subspaces
of moduli space as well.
Unlike their 1/2 BPS counterparts however, these
subspaces are unbounded,
and in fact divide the moduli space into two regions.
As we will show using the string web picture,
the 1/4 BPS states exist only in
one of the regions, and must therefore decay on the
marginal stability subspace.

Consider what happens to the 3-string junction
(figure~1a), with D3-brane boundaries,
as one of the external strings degenerates,
{\it i.e.} as one of the
D3-branes moves toward the junction point
(figure~5a,b). When it coincides
with the junction,
the string ending on it has vanishing length,
and we are left with two open strings, which can now
separate along the common
D3-brane. If we continue to move the D3-brane in the
same direction, we find
that a BPS 3-string junction no longer exists, because
the angle conditions (\ref{anglecondition}) cannot be
satisfied. We conclude that the corresponding 1/4 BPS state
$((1,0),(0,1))$ decays into two 1/2 BPS states,
{\it e.g.} $((1,1),(0,0)) + ((0,-1),(0,1))$,
corresponding to the two open strings, when the
D3-brane coincides with the
junction point. The loci of the junction point allowed
by supersymmetry defines a
marginal stability subspace for the 1/4 BPS state in question.
The projection of
this subspace on the plane of the string web gives a curve.
Since supersymmetry
fixes the angles between the strings, this curve is simply a
circular arc between the ends of the other two strings,
with a radius given by (label the two non-shrinking strings
$a,b$)
\be
  r = {|\vec{R}_a - \vec{R}_b|\over
    2\sin\theta_{ab}} \; .
\ee
By reflection symmetry, there is another arc on the other
side. This gives a
closed curve, which divides moduli space into a region
inside the curve and a
region outside the curve. The 1/4 BPS state exists only
outside.
This is reminiscent of the behavior of BPS states in
Seiberg-Witten theory \cite{sw}. One
could similarly consider shrinking one of the other
two strings. This leads to
two other curves of marginal stability.

This result generalizes to all string webs. For
example, for the 1/4 BPS states arising from the string
web in figure~1d there are two marginal stability curves
for each external string that shrinks
(figure~5c,d,e).
\begin{figure}[htb]
\epsfxsize=5in
\centerline{\epsffile{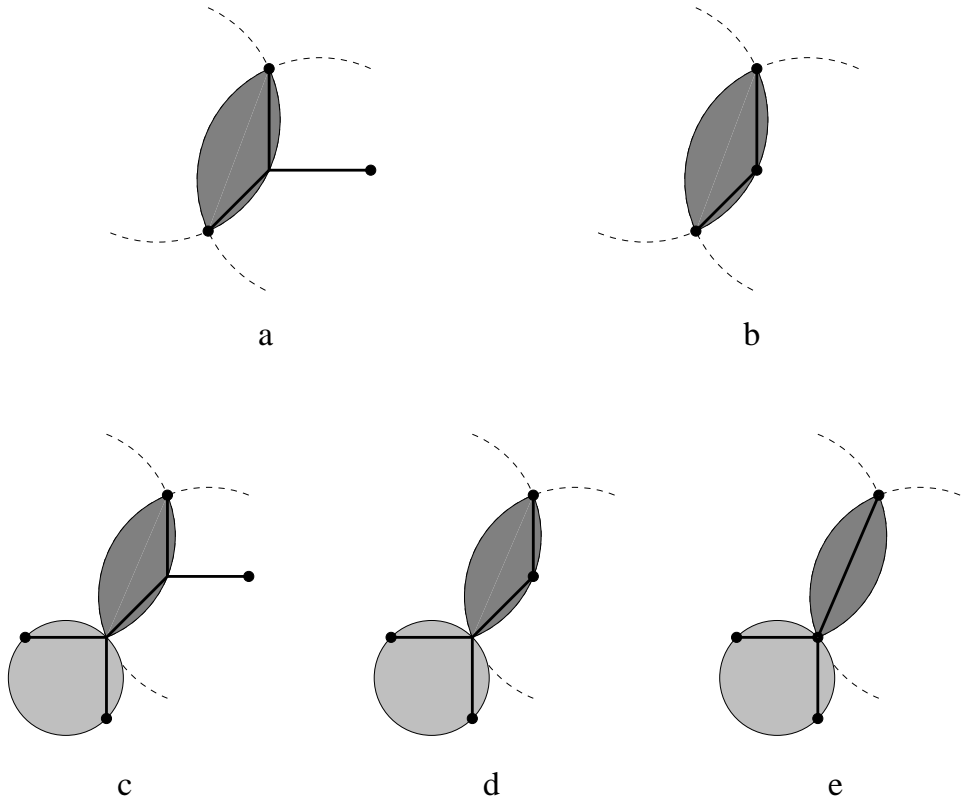}}
\caption{Curves of marginal stability and decay of 1/4 BPS
states. (a) 3-string junction decays into (b) two strings; 
(c) string web of fig.~1d decays first into (b) a string
and a 3-string junction, which proceeds to deacy into (c) two
more strings.}
\end{figure}

\section{Long BPS states and stable non-BPS states}
\label{unprotected_BPS}
\setcounter{equation}{0}

\noindent We have seen that as one moves in moduli space
one can encounter a wall of marginal stability where BPS states seem either to
decay, as in the case of 1/4 BPS states in $SU(3)$, or not, as in the case of
1/2 BPS states in $SU(3)$. We shall now see that there is yet another effect
that can occur, namely that a BPS state can become non-BPS, but remain stable.
The simplest example of this occurs in $SU(4)$.

Consider a planar string web with 4 or more external strings, for example
figure~1d. The moduli are a set of $N_c-1$ vectors in $\IR ^6$. For $N_c=3$ they
are always coplanar. For $N_c=4$, on the other hand, planar configurations occur
only on a submanifold of co-dimension 4.

 From section~3 we know that in
planar configurations supersymmetric webs exist for any
given charges and any
$N_c$, as long as curves of marginal stability are not
crossed. In particular,
the state in figure~1d will remain supersymmetric if we
make small {\it planar}
changes in the moduli.

Consider moving one of the D3-branes off the plane.
Intuition about ropes
suggests that the web deforms in such a way as
to minimize its mass. Because
each junction would be at mechanical equilibrium,
the three strings that meet
there should lie in one plane. Once the configuration is
non-planar, the planes
corresponding to the two junctions will no longer be
parallel, and the
configuration will break supersymmetry. For example,
in figure~1d we would get
two non-parallel fundamental strings, which break all
the supersymmetries.

This would seem to violate the stability of BPS states against changes in the
moduli. Such stability is guaranteed for short (or medium) multiplets, since
they cannot be smoothly connected to non-BPS states which come in long
multiplets. The resolution is that the aforementioned BPS state is {\it not} in
a medium representation. Indeed, from (\ref{n_FZM}) we see
that there are 16 FZM, only 12 of which correspond to 
broken supersymmetries. 
This means that the BPS state transforms as a {\it long}
multiplet, and is therefore not
protected against becoming non-BPS. We may call these states ``long
BPS multiplets" or ``accidental BPS", as they occur only on the submanifold of
planar configurations.

The long BPS state still satisfies the exact mass
formula $M=Z_+$, since
it is an algebraic consequence of unbroken supersymmetry. The non-planar webs
are not BPS, but we may still look for {\it extremal}
states of minimal mass
$M_{ex}$ with the given charges. The classical mass of a non-planar web can be
computed, and is found to be larger than the BPS bound. Furthermore, this mass
may receive quantum corrections. If we assume that these states survive the
transition to the quantum theory, and that BPS states are still absent, we would
have a strict inequality
\be
M_{ex}>Z_+ \; .
\label{extreme}
\ee
One can choose some parameter of deviation from planarity $\phi$,
and expand both
quantities as a function of $\phi$ (assuming the function
is regular at $\phi
=0)$,
\bea
 M_{ex} &=& m_0+m_1
\phi + m_2\phi ^2 /2 + \cdots \nonumber\\
 Z_+ &=& m_0+z_1
\phi + z_2\phi ^2 /2 + \cdots \; ,
\eea
where $m_i$ are the full non-perturbative coefficients of the
mass of the
non-BPS state. From (\ref{extreme}) we conclude that
\bea
m_1 & = & z_1 \nonumber \\
 m_2 & \geq & z_2 \; .
\label{masscoefs}
\eea
Since the planar
configuration is invariant under reflections,
$\phi$ may be chosen so that
$M_{ex},Z_+$ are even functions of $\phi$,
and (\ref{masscoefs}) is simplified.

\begin{center}
\large{ACKNOWLEDGMENTS}
\end{center}
OB thanks Matthias Gaberdiel and Barton Zwiebach for helpful discussions. BK
would like to thank Lenny Susskind; the hospitality at Tel-Aviv University and
in particular Andreas Brandhuber, Jacob Sonnenschein and Shimon Yankielowicz;
the hospitality of the Institute for Advanced Study, and in particular Amihay
Hanany, Dan Kabat, Nathan Seiberg, Savdeep Sethi and Edward Witten; the
hospitality of the workshop at the ITP, Santa Barbara and in particular Hirosi
Ooguri; and discussions with Renata Kallosh, Arvind Rajaraman and Scott Thomas.
OB is supported in part by the NSF under grant PHY-92-18167. BK is supported by
NSF grant PHY-9219345.

\appendix
\section{String webs and the BPS formula}
\setcounter{equation}{0}

\noindent We give here an alternative proof of the
equality of the mass of a web and the BPS formula. Consider the 4d $N=4$ gauge
theory, with $SU(N_c)$ gauge group and a complex coupling constant $\tau$. Its
moduli space is parameterized by $N_c$ 6-vectors $\vec{\Phi}_a=\Phi_a^I$, which
satisfy $\sum_{a=1}^{N_c}{\vec{\Phi}_a}=0$, and which generically break the
gauge group to $U(1)^{N_c-1}$. Consider a state with electric and magnetic
charges $(p_a,q_a)$ under the various $U(1)$'s, where $\sum{p_a}=\sum{q_a}=0$.
This state can be realized as a string web with external strings carrying
charges $(p_a,q_a)$.

Define the 2 vectors
$q^A_a=(\mbox{Re}(p_a+\tau q_a), \mbox{Im}(p_a+\tau q_a))$.
The central charge is determined in terms of a $2\times 6$ matrix
\be
Z^{AI}=\sum_{a}{q^A_a \Phi^I_a} \; .
\label{Z_AJ}
\ee
Consider first a 3-string web with no internal faces.
Later we shall
generalize to an arbitrary web. Choose the directions
of the $\Phi$ coordinates
in the plane of the web such that
$\Delta(\Phi^1+i\Phi^2)_a=(p_a+\tau q_a)$, as
required by supersymmetry.
We see that the charge vectors $q^A_a$ are
parallel to the plane vectors $\Phi^I_a$:
\be
\Phi^I_a=r_a q^I_a \qquad (I=1,2) \;,
\ee
where $r_a$ are real and non-negative.
The origin of the plane is
chosen at the junction. The central charge matrix is now given by
\be
Z^{AI}=\sum_{a}{q^A_a \Phi^I_a}=
  \sum_{a}{r_a q^A_a q^I_a} \; ,
\ee
which is a {\it symmetric} $2\times 2$ matrix. In this case the formula
(\ref{Zpm}) for the central charge simplifies to
\be
Z_+=\left| tr(Z) \right|,
\ee
assuming $det(Z) \geq 0$, which holds in our case.
Computing $tr(Z)$ we get
\bea
tr(Z)&=& \sum_{i,I}r_i \cdot q^I_i q^I_i \nonumber \\
     &=& \sum_{i} \left| \vec{q} \right|
 \left| \vec{\Phi} \right| =m_{web} \; .
\eea
After observing that if a web exists its mass equals the
central charge, let us
show that such a web can always be found, unless a marginal
stability wall is crossed. To show that we will consider an
existing web, which is not at
marginal stability, and we will show that for any
small change in the location
of one of the D3-branes a web still exists.
It is enough to consider a single D3-brane.
If we move it along the string that is attached to it,
no change is
required in the web. If we move the D3-brane transverse
to the string attached
to it, we have one rotational degree of freedom that
allows us to solve for the
web. This is the angle defining the
directions of the
$\Phi^{1,2}$ axes.

To generalize for webs of arbitrary $N_c$ with
possible internal faces,
we first use the bosonic zero modes to shrink all
internal faces. We are left
with a ``tree'' like web. We set the origin at one
of the vertices, and
transform (\ref{Z_AJ}) from a sum over external strings,
to a sum over all strings
\be
Z^{AI}=\sum_{a=1}^{E_{ext}} q^A_a \Phi^I_a =
\sum_{a=1}^{E_{ext}} q^A_a \sum_{i\in a}{\Phi^I_i} =
  \sum_{i=1}^E q^A_i \Phi^I_i \; ,
\ee
where the notation $i \in a$ means that the string $i$
is on the route from the
origin to the external string $a$, and it is well defined
for a tree web. Charge
conservation at vertices is used in the last equality.
At this point the
previous argument can be repeated.

\begingroup\raggedright
\endgroup

\end{document}